\documentclass[aps,superscriptaddress,showkeys,showpacs,longbibliography,nofootinbib]{revtex4-2}
\usepackage{graphicx}
\usepackage{bm}
\usepackage{graphicx}
\usepackage{amsmath}
\usepackage{mathrsfs}
\usepackage{ulem}
\usepackage{color}
\usepackage[colorlinks, linkcolor=red,anchorcolor=green,citecolor=blue]{hyperref}

\newcommand{\nn}{\nonumber}
\newcommand\diag{\operatorname{diag}}

\graphicspath{{./figs/}}

\begin{document}  
	
\title{The linear mode analysis and spin relaxation}

\author{Jin Hu}
\email{hu-j17@mails.tsinghua.edu.cn}
\affiliation{Department of Physics, Tsinghua University, Beijing 100084, China}

\begin{abstract}
In this paper, a detailed analysis on normal modes of the linearized Hermite collision operator is presented, which follows from  linearizing spin Boltzmann equation for massive fermions proposed in \cite{Weickgenannt:2021cuo} with the non-diagonal  part of the transition rate neglected and approximating what we got with a mutilated operator.  With the assumption of  total angular momentum conservation, the collision term is proved to well describe the equilibrium state and gives proper interpretation for collisional invariants, thus is relevant for the research on local spin polarization.  Following the familiar fashion as used in quantum mechanics, we treat the problem of solving normal modes as a degenerate perturbation problem and calculate the dispersion relations for intriguing eleven zero modes, which form one-to-one correspondence to all collisional invariants. We find that the results of spinless modes appearing in ordinary hydrodynamics  are consistent with available conclusions in  textbooks. As for spin-related modes, we obtain the frequencies up to second order in wave vector and relate them with the dissipation of spin density fluctuation. In addition, the ratio of two relaxation time scales for spin and momentum is shown as a function of reduced mass, which reads that based on present framework spin equilibration is almost as slow as momentum equilibration as far as the strange quark spins in quark gluon plasma (QGP) are concerned. 
\end{abstract}




\maketitle	
\section{Introduction}
Recent developments in the experiments of relativistic non-central heavy-ion collisions have seen great progress in measuring the net spin polarization of $ \Lambda $ and $\bar{\Lambda}$ hyperons \cite{STAR:2017ckg,Alpatov:2020iev}. In the early stage of evolution of the hot, dense matter produced in non-central heavy-ion collisions, the medium carries a huge total angular momentum that is converted into spin angular momentum of the particles of final states via spin-orbit coupling.  Theoretical researches on the global polarization of $ \Lambda $ hyperons have long attracted extensive interests  and correspondent numerical results  satisfy experimental data well \cite{Wei:2018zfb,Karpenko:2016jyx,Csernai:2018yok,Li:2017slc,Bzdak:2017shg,Shi:2017wpk,Sun:2017xhx,Ivanov:2019wzg,Xie:2017upb}. However, they have difficulties in providing satisfying explanations for the measurements of differential spin polarization, i.e. the dependence of $ \Lambda $ polarization  on the azimuthal angle and transverse momentum \cite{Adam:2019srw,Adam:2018ivw}, which is usually called ``spin sign problem'' \cite{Becattini:2017gcx,Xia:2018tes}. Resolving this problem calls for new theoretical frameworks and concerns one still unsettled question of  how strange quark spin comes to equilibrium. There are two possible mechanisms by which strange quark spin could equilibrate. One is that fluctuations of the vorticity will drive the spins towards equilibrium just as fluctuations around an external magnetic field do, the other is dynamic mechanism originating from the scatterings between strange quarks and other particles within the medium \cite{Kapusta:2019ktm}. If taking an analogy with transport equation, the proposed mechanisms exactly refer to the external field term and the collision term. We here address that the crucial point for answering these questions lies in the appropriate extraction of spin equilibration time scale. In short, new framework must take into account the influence of dissipation instead of treating spin as a independent equilibrated quantity.  Among all the candidates, spin hydrodynamics and spin transport  are thought to be  promising. Spin hydrodynamics extends the description of ordinary fluid theory  by including  spin degree of freedom, based on which the relaxation of spin and the evolution of vorticity become the focus of theoretical research, while spin transport, namely, focuses on constructing a consistent theory for both spin and momentum transport. Though, in principle, the latter possesses a wider range of application, the constructed transport equation on the market is too involved to be solved on account of the complexity and non-linearity.

There are many developments in the investigation of  spin hydrodynamic. Among them, ``ideal'' spin hydrodynamics was proposed in the context of the QGP~\cite{Florkowski:2017ruc} and  for massive spin-$1/2$ fermions \cite{Peng:2021ago}.  Recently, first-order spin hydrodynamics including non-equilibrium corrections has also been put into efforts~\cite{Hattori:2019lfp,Bhadury:2020cop,Fukushima:2020ucl,Hu:2021pwh,Hu:2021lnx}. 
 In a recent work, the authors construct a second-order spin hydrodynamic theory based on the method of  moment expansion \cite{Weickgenannt:2022zxs} from spin transport equation \cite{Weickgenannt:2021cuo,Weickgenannt:2020aaf}, which is also our starting point for linear mode analysis.  As a reminder, one can also see   \cite{Yang:2020hri,Wang:2020pej,Sheng:2021kfc,Chen:2021azy,Lin:2021mvw,Wang:2021qnt,Yang:2021fea}  for continuous efforts on the research  of spin transport theory. 

In this paper, we present a detailed analysis on normal modes of the linearized  collision operator. To that end,  we adopt the transport equation along with a nonlocal collision kernel proposed in \cite{Weickgenannt:2021cuo} as the start point of our calculation. The linear mode analysis is closely associated with hydrodynamics  because the theory of fluids can be completely constructed from these normal modes, which are nothing but collisional invariants, i.e,  the microscopic correspondence of macroscopic conserved laws.  
Moreover, one can relate dispersion relations of normal modes with the dissipation of various fluctuation amplitudes, from which various relaxation time scales can be extracted. As is raised in previous paragraphs, the comparison for these time scales is significant for investigating spin polarization and  spin equilibration.
This paper is organized as follows. In Sec.~\ref{secqu}
we present a short review of spin Boltzmann equation with a nonlocal collision term. In Sec.~\ref{eq}  the equilibrium distribution function is briefly discussed and the conditions for global equilibrium are also obtained. After that,  the mutilated operator inheriting the important properties of full linearized collision operator  is proposed in Sec.~\ref{lated}.  In Sec.~\ref{expansion}, we present the detailed analysis on normal modes of the approximated linearized operator, i.e., mutilated collision operator following the fashion as used in quantum mechanics \cite{Balescu, Resi}. Summary and outlook are given in Sec.~\ref{su}. Natural units $k_B=c=\hbar=1$ are utilized. The metric tensor here is given by $g^{\mu\nu}=\diag(1,-1,-1,-1)$ , while $\Delta^{\mu\nu} \equiv g^{\mu\nu}-u^\mu u^\nu$ is the projection tensor orthogonal to the four-vector fluid velocity $u^\mu$.

In addition, we employ the symmetric/antisymmetric shorthand notations:
\begin{eqnarray}
X^{( \mu\nu ) } &\equiv& (X^{ \mu\nu } + X^{ \nu \mu})/2, \\
X^{[ \mu\nu ] } &\equiv& (X^{ \mu\nu } - X^{ \nu \mu})/2, \\
X^{\langle \mu\nu \rangle}&\equiv&
	\bigg(\frac{\Delta^{\mu}_{\alpha} \Delta^{\nu}_{\beta} 
		+ \Delta^{\nu}_{\alpha} \Delta^{\mu}_{\beta}}{2}
	- \frac{\Delta^{\mu\nu} \Delta_{\alpha\beta}}{3}\bigg)X^{\alpha\beta}.
\end{eqnarray}
Specially, we decompose the derivative $\partial$ according to 
\begin{align}
\partial^\mu=u^\mu D+\nabla^\mu,\quad D\equiv u^\mu\partial_\mu,\quad \nabla^\mu\equiv\Delta^{\mu\nu}\partial_\nu.
\end{align}

\section{Review of the nonlocal transport equation }
\label{secqu}	
We start with the spin Boltzmann equation with a nonlocal collision term for massive fermions  proposed in \cite{Weickgenannt:2021cuo}.
Assuming that the evolution of the system of our interest is governed by the proposed on-shell Boltzmann equation, which extends the  phase space to incorporate the variable $\bm{s}$ as a classical description of spin degrees of freedom,
\begin{align}
\label{boltz}
p\cdot \partial f(x,p,\bm{s})=C[f],
\end{align}	 
with  the nonlocal collision term
\begin{eqnarray}
\label{cf}
C[f] &  \equiv& \int d\Gamma_1 d\Gamma_2 d\Gamma^\prime\,    
\mathcal{W}\,  
[f(x+\Delta_1,p_1,\bm{s}_1)f(x+\Delta_2,p_2,\bm{s}_2)
-f(x+\Delta,p,\bm{s})f(x+\Delta^\prime,p^\prime,\bm{s}^\prime)],
\end{eqnarray}
where the measure appearing in the collision kernel  is defined as $d\Gamma \equiv d^4p\, \delta(p^2 - m^2)dS(p)$ , the newly introduced measure $dS(p)$ is given in Eq.(\ref{dsp}) and the other collision term corresponding to only  spin changes in \cite{Weickgenannt:2021cuo} is neglected.
 Here we note the transition rate $\mathcal{W}$ is 
\begin{eqnarray}
\mathcal{W}&\equiv& \delta^{(4)}(p+p^\prime-p_1-p_2)
\frac{1}{8} \sum_{s,r,s',r',s_1,s_2,r_1,r_2} h_{s r} (p,\bm{s})h_{s^\prime r^\prime}(p^\prime, \bm{s}^\prime) \,  
h_{s_1 r_1}(p_1, \bm{s}_1)\, h_{s_2 r_2}(p_2, \bm{s}_2) \nn \\
&&\times \langle{p,p^\prime;r,r^\prime|t|p_1,p_2;s_1,s_2}\rangle
\langle{p_1,p_2;r_1,r_2|t^\dagger|p,p^\prime;s,s^\prime}\rangle \;
\label{local_col_GLW_after}
\end{eqnarray}
with 
\begin{equation}
\label{hsr}
h_{s r}(p,\bm{s}) \equiv \delta_{s r}+  \frac{1}{2m}\, \bar{u}_s(p)\gamma^5\bm{s} \cdot\gamma u_r(p)\; .
\end{equation} 
As is shown by the two terms in Eq.(\ref{hsr}), we split $\mathcal{W}$ into the unpolarized part and polarized one. Namely, when neglecting the non-diagonal part in Eq.(\ref{hsr}), i.e., the term linearized to spin $\bm{s}$, the transition rate takes exactly  the unpolarized  form (sum up the final states and average the initial states),
\begin{eqnarray}
\bar{\mathcal{W}}&\equiv& \delta^{(4)}(p+p^\prime-p_1-p_2)\,
\frac{1}{8} \sum_{r,r',r_1,r_2} \langle{p,p^\prime;r,r^\prime|t|p_1,p_2;r_1,r_2}\rangle
\langle{p_1,p_2;r_1,r_2|t^\dagger|p,p^\prime;r,r^\prime}\rangle \;,
\label{local_col_GLW_after2}
\end{eqnarray}
where the $\gamma$ matrices, spinor $u_s(p)$ and spin indices $r,s$  correspond to the spinor description for fermions as often used, and the matrix element of $t$ is the conventional scattering amplitude defined in quantum field theory. To proceed, we comment that the crucial point for the nontrivial extension of the collision term lies in the spatial shift $\Delta$ manifesting the nonlocality of the collisions,
\begin{equation}
\label{deltanon}
\Delta^\mu\equiv -\frac{1}{2m(p\cdot\hat{t}+m)}\, 
\epsilon^{\mu\nu\alpha\beta}p_\nu \hat{t}_\alpha \bm{s}_{\beta}\;,
\end{equation}
where $\hat{t}^\mu$ is the time-like unit vector which is $(1,\boldsymbol{0})$ 
in the frame where $p^\mu$ is measured.
 The  collision shift well captures the properties of spin-orbit coupling in nonlocal collisions, thus is highly relevant for discussing the spin sign problem of local polarization of $\Lambda$.

 To proceed, we move on to  the classical spin. Spin here is treated as an additional variable in phase space
 \cite{Zamanian:2010zz,Ekman:2017kxi,Ekman:2019vrv,Florkowski:2018fap,Bhadury:2020puc,Weickgenannt:2020aaf}, 
 which  immediately connects the first-principle 
 quantum description to a ``classical'' description of spin. 
  In previous paragraphs, we have introduced  the covariant
  integration measure for spin 
  \begin{equation}
  \label{dsp}
  \int dS(p)  \equiv \sqrt{\frac{p^2}{3 \pi^2}} \int d^4\bm{s}\, \delta(\bm{s}\cdot\bm{s}+3)
  \delta(p\cdot \bm{s})\;.
  \end{equation}
Then the following useful integrals can be easily obtained via rather straightforward calculations,
  \begin{subequations}
  	\begin{eqnarray}
  	\label{f1}
  	\int dS(p) & = & 2\;, \\
  	\label{f2}
  	\int dS(p)\, \bm{s}^\mu & = & 0 \;, \\
  	\label{f3}
  	\int dS(p)\, \bm{s}^\mu \bm{s}^\nu & = & - 2 \left( g^{\mu \nu} - \frac{p^\mu p^\nu}{p^2} \right)\;,\\
  	  	\label{f31}
  	  	\int dS(p)\,\bm{s}^\lambda \bm{s}^\mu \bm{s}^\nu & = &0,\\
  	\label{f4}
\int dS(p) \Sigma^{\mu\nu}_{\bm{s}}\Sigma_{\bm{s}}^{\alpha\beta}
&=&\frac{2}{m^2}(g^{\mu\alpha}g^{\nu\beta}p^2+g^{\mu\beta}p^{\nu}p^{\alpha}+g^{\nu\alpha}p^{\mu}p^{\beta}-[\mu\leftrightarrow\nu])\;,
  	\end{eqnarray}  	
  \end{subequations}
    where the dipole tensor $\Sigma_{\bm{s}}^{\mu\nu}\equiv -\frac1m \epsilon^{\mu\nu\alpha\beta}p_\alpha \bm{s}_\beta$ is interpreted as the spin angular momentum of the particle \cite{Weickgenannt:2021cuo}.
    
With the extended phase space,  our interesting tensors such as particle current, energy-momentum tensor and spin tensor  can be convenient written as
 \begin{align}
 \label{N}
 &N^{\mu}\equiv\int\,d\Gamma \,p^\mu f(x,p,\bm{s}),\\
 \label{T}
 &T_{\text{HW}}^{\mu\nu}\equiv\int \,d\Gamma \,p^\mu p^\nu f(x,p,\bm{s})+T_{\text{HW}}^{[\mu\nu]},\\
 \label{S}
 &S_{\text{HW}}^{\lambda,\mu\nu}\equiv\int \,d\Gamma \,p^\lambda (\frac{1}{2}\Sigma^{\mu\nu}_{\bm{s}}-\frac{1}{2m^2}p^{[\mu}\partial^{\nu]} )f(x,p,\bm{s}),
 \end{align}
 where we have chosen the psudo-gauge proposed by Hilgevoord and Wouthuysen (HW) \cite{HILGEVOORD19631,hilgevoord1965covariant,Weickgenannt:2020aaf}, the spin tensor of which is proved to be conserved in global equilibrium and non-conserved away from global equilibrium because of mutual conversion between spin and orbital angular momentum  and thus more intuitively suitable for our discussion herein and we keep $T_{\text{HW}}^{[\mu\nu]}$ to stress that $T^{\text{HW}}$ is not symmetric and  is in the second order in $\hbar$ expansion  while its form is left unspecified within our paper, see \cite{Weickgenannt:2022zxs,Weickgenannt:2020aaf} for more details. 

 
   By taking Landau choice of fluid velocity and Landau matching conditions, 
  \begin{align}
 T_{\text{S}}^{\mu\nu}u_\nu=eu^\mu, \quad u_\mu N^\mu=u_\mu N^\mu_{\text{eq}}, \quad u_\mu T_{\text{S}}^{\mu\nu}u_\nu=u_\mu T^{\mu\nu}_{\text{eq}}u_\nu,\quad  u_\lambda J^{\lambda,\mu\nu}=u_\lambda J_{\text{eq}}^{\lambda,\mu\nu},
  \end{align}
  where $T_{\text{S}}^{\mu\nu}$ refers to the symmetric part of $T^{\mu\nu}_{\text{HW}}$ and the total angular momentum tensor is defined as $J^{\lambda,\mu\nu}\equiv S_{\text{HW}}^{\lambda,\mu\nu}+x^\mu T_{\text{HW}}^{\lambda\nu}-x^\nu T_{\text{HW}}^{\lambda\mu}$,
   we allow the following decompositions,
   \begin{align}
   \label{N1}
   &N^{\mu}=nu^\mu+V^\mu,\\
   \label{T1}
   &T_{\text{S}}^{\mu\nu}=eu^\mu u^\nu-P\Delta^{\mu\nu}+\pi^{\mu\nu}+\Pi\Delta^{\mu\nu},\\
   \label{S1}
   &S_{\text{HW}}^{\lambda,\mu\nu}=u^\lambda S^{\mu\nu}+\delta S^{\lambda,\mu\nu},
   \end{align}
   where $n, e, P, S^{\mu\nu}$ are the particle number density, energy density, static pressure and spin density, and the dissipative quantities $V^\mu, \pi^{\mu\nu}$ and $\Pi$ are the diffusion current, shear stress tensor, bulk viscous pressure respectively. Note that generally $u_\lambda \delta S^{\lambda,\mu\nu}\neq 0$.

 \section{Equilibrium}
  \label{eq}	
 In this section, we will show that the collision term Eq.(\ref{cf}) is consistent with the standard form of spin-dependent local equilibrium distribution function \cite{Becattini:2013fla,Florkowski:2017ruc},
 \begin{align}
 \label{feq}
 &f_{\text{leq}}(x,p,\bm{s})=\frac{1}{(2\pi)^3}\exp[\xi-\beta\cdot p+\frac{\Omega_{\mu\nu}\Sigma_{\bm{s}}^{\mu\nu}}{4}], 
 \end{align}
 where $\Omega_{\mu\nu}$ represents spin potential, while $\beta^\mu\equiv\frac{u^\mu}{T},\xi\equiv \frac{\mu}{T},\beta\equiv\frac{1}{T}$ with the temperature $T$,  and  the chemical potential $\mu$  introduced for conserved particle number (only elastic scatterings are considered). The exponent in Eq.(\ref{feq}) is exactly the linear combination of all conserved quantities, and $\xi,\beta$ and $\Omega^{\mu\nu}$ are the correspondent Lagrangian multipliers maximizing the total entropy of the system. To prove this, the substitution of Eq.(\ref{feq}) into  Eq.(\ref{cf}) leads to
\begin{eqnarray}
\label{cf00}
C[f_{\text{leq}}] 
&= & -\frac{1}{(2\pi)^6} \int d\Gamma^\prime d\Gamma_1 d\Gamma_2  \,
\bar{\mathcal{W}}\, F[p,p^\prime,p_1,p_2;\bm{s},\bm{s}^\prime,\bm{s}_1,\bm{s}_2] \exp(2\xi-\beta\cdot(p+p^\prime))\nn\\
& &\times  \big[-\partial_\mu\xi \frac{}{}
\left(\Delta_1^\mu +\Delta_2^\mu -\Delta^\mu -\Delta^{\prime\mu}
\right)+\partial_\mu\beta_\nu \frac{}{}
\left(\Delta_1^\mu p_1^\nu+\Delta_2^\mu p_2^\nu-\Delta^\mu p^\nu-\Delta^{\prime\mu}
p^{\prime\nu} \right)\nn\\
&&
\quad\,\,- \frac 1 4\Omega_{\mu\nu}\left(\Sigma_{\bm{s}_1}^{\mu\nu}
+\Sigma_{\bm{s}_2}^{\mu\nu}
-\Sigma_{\bm{s}}^{\mu\nu}-\Sigma_{\bm{s}^\prime}^{\mu\nu}\right) \big],
\label{colleqq}
\end{eqnarray}
 where the local equilibrium distribution is Taylor expanded to first order in $\Omega$ assuming small spin potential (\,if the system in discussion is close to the state of global equilibrium, $\Omega$ is about the order of the gradient of $\beta$ field\,).

When including spin, it is the global equilibrium distribution function that makes the collision term vanish as long as the total angular momentum 
 \begin{align}
 J^{\mu\nu}=2\Delta^{[\mu}p^{\nu]}+\frac{1}{2}\Sigma^{\mu\nu}_{\bm{s}}
 \end{align}
 is conserved in a collision, which is distinct from traditional definition for local equilibrium. In that case, the conditions for vanishing Eq.(\ref{colleqq}) are
\begin{align}
\label{condition}
&\partial_{(\mu}\beta_{\nu)}=0,\quad \xi=\text{const},\nn\\
&\Omega_{\mu\nu}=-\partial_{[\mu}\beta_{\nu]}=\text{const},
\end{align}
which implies that  the spin potential $\Omega_{\mu\nu}$  is fixed to thermal vorticity $\frac{1}{2}(\partial_\nu \beta_\mu-\partial_\mu \beta_\nu)$.  $\beta^\mu$ can be further  decomposed into a translation ($a^\mu$) and a rigid rotation ($\Omega^{\mu\nu}x_\nu$)  in global equilibrium,
\begin{align}
&\beta^\mu=a^\mu+\Omega^{\mu\nu}x_\nu,\quad a^\mu=\text{const},
\end{align}
which are consistent with the previous conclusions drawn in  \cite{Becattini:2013fla,Florkowski:2017ruc}. 
One may observe that Eq.(\ref{cf00}) contains an extra dimensionless factor $F$ compared to Eq.(\ref{cf}) and $\mathcal{W}$ is replaced by $\bar{\mathcal{W}}$. The reason why these changes are necessary is elaborated in \cite{Hu:2022lpi} and we want to emphasize that the reformed collision operator or collision term respects hermity, non-negative property and detailed balance as well as the collisional invariance for total angular momentum, which is a necessary condition for a  consistent description of global equilibrium.

 \section{Mutilated collision operator}
 \label{lated}	

Following \cite{Hu:2022lpi}, we choose a quiescent (also unpolarized) background fluid i.e., $u^{\mu}=(1,0,0,0),\, \Omega^{\mu\nu}=0$ with $\delta u^\mu$ and $\delta\Omega^{\mu\nu}$ treated as perturbations. With more details given therein,
we start with the equation for $\tilde{\chi}$ (we want to find a solution of the form  $\chi\sim\tilde{\chi}e^{-ik\cdot x}$ to the linearized spin Boltzmann equation)
\begin{align}
\label{cf2}
&\tau \omega\tilde{\chi}+\hat{p}^\mu \kappa_\mu \tilde{\chi}+L_2[\tilde{\chi}]=-iL_1[\tilde{\chi}], 
\end{align}
with notations
\begin{align}
\label{pert}
\tau\equiv \frac{p\cdot u}{T},\quad \omega\equiv \frac{u\cdot k}{n\sigma(T)},\quad\hat{p}\equiv\frac{p}{T},\quad \kappa^\alpha\equiv\frac{\Delta^{\alpha\beta}k_\beta}{n\sigma(T)},\quad \kappa\equiv\sqrt{-\kappa\cdot\kappa},\quad l^\alpha\equiv\frac{\kappa^\alpha}{\kappa},
\end{align}
where  $\sigma(T)$ is an arbitrary constant with the dimension of cross sections \cite{DeGroot:1980dk}, $\chi$ denotes the deviation from the global equilibrium distribution, 
  and $L_1, L_2$ are the dimensionless collision operators  given by
\begin{eqnarray}\label{colleqqcl1}
L_1[\phi]
&\equiv &\frac{1}{(2\pi)^3n\sigma(T)T} \int d\Gamma^\prime d\Gamma_1 d\Gamma_2  \,
\bar{\mathcal{W}} \,  F[p,p^\prime,p_1,p_2;\bm{s},\bm{s}^\prime,\bm{s}_1,\bm{s}_2] \exp(\xi-\beta\cdot p^\prime\,)\nn\\
&\times & \Big[\phi(k,p,\bm{s})+\phi(k,p^\prime,\bm{s}^\prime)-\phi(k,p_1,\bm{s}_1)-\phi(k,p_2,\bm{s}_2)\,\Big],\\
L_2[\phi]
  &\equiv &\frac{1}{(2\pi)^3T} \int d\Gamma^\prime d\Gamma_1 d\Gamma_2  \,
  \bar{\mathcal{W}} \,   F[p,p^\prime,p_1,p_2;\bm{s},\bm{s}^\prime,\bm{s}_1,\bm{s}_2]\exp(\xi-\beta\cdot p^\prime\,)\nn\\
  & \times&  \Big[\Delta\cdot \kappa\phi(k,p,\bm{s})+\Delta^\prime\cdot \kappa\phi(k,p^\prime,\bm{s}^\prime) -\Delta_1\cdot  \kappa\phi(k,p_1,\bm{s}_1) -\Delta_2\cdot \kappa \phi(k,p_2,\bm{s}_2)  \Big].
  \label{colleqqs2}
  \end{eqnarray}
  Here we remind that the inner product is defined in global equilibrium  as
  \begin{align}
  \label{inner}
  (B,C)=\frac{1}{(2\pi)^3}\int d\Gamma \exp(\xi-\beta\cdot p)B(p,\bm{s}) C(p,\bm{s}).
  \end{align}

   According to the calculations in \cite{Hu:2022lpi}, $L_2$ contributes nothing to the dispersion law, which is known from the fact that $L_2$ is absent in  the formulas for $\omega$. Therefore, we neglect this term from now on but the complexity of solving Eq.(\ref{cf2}) does not decrease a lot. The linearized collision operator has a very complicated structure even for the simplest interaction. However, it seems reasonable that the qualitative features of the normal modes depend only upon the universal properties independent of which interaction to take, i.e., that the spectrum of the collision operator $-L_1$ is composed of a eleven-fold degenerate zero, which are nothing but all collisional invariants, and a sequence of negative eigenvalues. These zero normal modes are protected by the conservation laws or essentially the translational and Lorentz symmetries. 
   In the linear analysis, a small fluctuation $\chi$  on top of background distribution can always be expanded with a set of eigenstates of the linearized collision operator formally \cite{Balescu}
   \begin{align}
   \chi=\sum_{n=0}^{\infty}a_ne^{\gamma_n t}|\gamma_n\rangle,
   \end{align}
   where $\gamma_n$ and $|\gamma_n\rangle$ constitute the representation of eigen spectrum of linearized collision operator,  $a_n$ denotes the expansion coefficient and $t$ is seen as a dimensionless time. As the system evolves in time $t$, protected modes with $\gamma_n=0$ remain unchanged and other negative modes become less important after a characteristic time scale.
   If taking this idea as an ansatz, the full linearized operator can be approximated as a mutilated collision operator in which all the negative eigenvalues collapse into a single eigenvalue of infinite degeneracy \cite{deboer},
  \begin{align}
  \label{l1}
 -L_1\approx n\sigma u\cdot p\big(-\gamma+\gamma \sum_{n=1}^{11}|\lambda_n\rangle\langle \lambda_n|\big)
  \end{align}
  with $|\lambda_n\rangle$ being the eleven degenerate orthonormal eigenvectors of zero eigenvalue and $-\gamma$ is the remaining negative eigenvalue (it is suggestive that $-\gamma$ is chosen to be the largest one of all negative eigenvalues). To be concrete, we rewrite the Eq.(\ref{l1}) as 
  \begin{align}
  \label{appro}
  -L_1\phi(k,p,\bm{s})\approx -n\sigma u\cdot p\gamma\big(\phi-\sum_{n=1}^{11}(\tilde{\psi}_n,\tau\phi)\tilde{\psi}_n\;\big),
  \end{align}
  where the eigenfunction set is defined in Eq.(\ref{set}) and there are no differences between $\psi_n$ and $\tilde{\psi}_n$ for lack of the coordination $x$ dependence in both eigenfunction sets. It is easy to verify that this new defined operator (the right hand side of Eq.(\ref{appro})) has the basic features belonging to the full operator with respect to the orthonormal condition Eq.(\ref{one}). One may have doubt why taking the orthonormal condition with a weight function $\tau$. To answer this question, one should be informed  that the frequency $\omega$ of our interest is not the eigenvalue of the operator $L_1$ but $\tau\omega$ is viewing Eq.(\ref{unpert}). In addition, one may observe that the proposed approximation is exactly reformed relaxation time approximation (RTA) respecting $L_1|\lambda_n\rangle=0, n=1,2,\cdots 11$ and $-L_1|\lambda_n\rangle=-\gamma|\lambda_n\rangle$ with $n>11$. This extra factor $\tau$ is consistent with the form of relativistic relaxation time approximation. Last but not the least, this novel version of RTA is proved to reconcile the momentum dependence of the relaxation time with the conservation  laws. 
   In fact, when the relaxation time has momentum dependence as it should do in general, the argument via matching conditions used to resolve this contradiction also fails. Here we do not go into details further and similar results are also reported in a recent letter \cite{Rocha:2021zcw}. For simplicity, we require that $\gamma$ is  momentum independent hereafter and when nothing confusing happens, $L_1$  represents the mutilated operator or RTA operator instead of the full operator. Eventually the equation to be solved is
  \begin{align}
  \label{cf5}
  &\tau \omega\tilde{\chi}+\hat{p}^\mu \kappa_\mu \tilde{\chi}=-i\gamma\tau\big(\tilde{\chi}-\sum_{n=1}^{11}(\tilde{\psi}_n,\tau\tilde{\chi})\tilde{\psi}_n\;\big).
  \end{align}
  
   \section{Degenerate perturbation theory and linear mode analysis}
   \label{expansion}
  
 As a well-posed problem in the perturbation theory,  the solutions to Eq.(\ref{cf2}) can be sought in the following steps. Firstly, treat the gradients of the thermodynamic variables as a perturbation with respect to the linearized collision operator $-iL_1$, then we obtain an eigenvalue problem for the unperturbed equation,
\begin{align}
\label{unpert}
&-iL_1\tilde{\chi}^{(0)}= \tau\omega^{(0)}\tilde{\chi}^{(0)}.
\end{align}
Since  the mutilated operator is constructed based on the priori knowledge about the spectrum of the full collision operator, thus the conservation laws for energy-momentum, particle number and total angular momentum have already been contained and the eigenfunctions are all the linear combination of all collisional invariants $1, p^\mu$ and $J^{\mu\nu}$. Compared to ordinary hydrodynamic description, the six new modes arise from the nontrivial dynamics of spin and we identify them with the spin modes. Due to the eleven-fold degeneracy, the perturbation term should be taken into account to remove or partly remove the degeneracy.

For the first-order perturbation of $p\cdot \kappa$, the evaluation of the eigen spectrum can be done in a familiar fashion used in quantum mechanics. 	To proceed, we denote the $n$-th order eigenvalues and eigenfunctions as $\omega^{(n)}$ and $\tilde{\chi}^{(n)}$ for concreteness
\begin{align}
&\tilde{\chi}=\tilde{\chi}^{(0)}+\tilde{\chi}^{(1)}+\cdots,\nn\\
&\omega=\omega^{(0)}+\omega^{(1)}+\omega^{(2)}+\cdots.
\end{align}
 Following the procedures in quantum mechanics, $\tilde{\chi}^{n}$ with $n>0$ is  chosen as the combination of the eigenfunctions of Eq.(\ref{unpert}) excluding the zeroth-order eigenfunctions. In  order to break the eleven-fold degeneracy, the standard method of Schmidt orthogonalization is adopted
and the eigenfunctions  for these eleven-fold degenerate zeros  can be taken as
\begin{align}
\label{set}
&\tilde{\psi}_{1}=\frac{1}{\sqrt{V_{1,1}}}, \quad \tilde{\psi}_{2}=\beta\frac{u\cdot p-\frac{e}{n}}{\sqrt{V_{2,2}}},\quad \tilde{\psi}_{3}=\frac{\beta l\cdot p}{\sqrt{V_{3,3}}},\quad \tilde{\psi}_{4}=\frac{\beta j\cdot p}{\sqrt{V_{3,3}}}, \quad \tilde{\psi}_{5}=\frac{\beta v\cdot p}{\sqrt{V_{3,3}}},\nn\\
&\tilde{\psi}_{6}=\frac{u_\mu J^{\mu\nu}l_\nu}{\sqrt{V_{6,6}}},\quad \tilde{\psi}_{7}=\frac{u_\mu J^{\mu\nu}j_\nu}{\sqrt{V_{6,6}}},\quad \tilde{\psi}_{8}=\frac{u_\mu J^{\mu\nu}v_\nu}{\sqrt{V_{6,6}}},\quad \tilde{\psi}_{9}=\frac{l_\mu J^{\mu\nu}j_\nu}{\sqrt{V_{9,9}}},\nn\\
&\tilde{\psi}_{10}=\frac{l_\mu J^{\mu\nu}v_\nu}{\sqrt{V_{9,9}}}, \quad \tilde{\psi}_{11}=\frac{j_\mu J^{\mu\nu}v_\nu}{\sqrt{V_{9,9}}},
\end{align}
to fulfill the orthonormal condition
\begin{align}
\label{one}
(\tilde{\psi}_\alpha,\tau \tilde{\psi}_\beta)=\delta_{\alpha\beta},
\end{align}
where the definitions of two auxiliary vectors $j, v$ and the normalized factor $V_{i,j}$ are all put in Appendix.(\ref{int1}).
We now seek the solutions to the inhomogeneous integral equation for $\tilde{\chi}^{(1)}$ 
\begin{align}
\label{pert1}
&-iL_1\tilde{\chi}_\alpha^{(1)}= \tau\omega_\alpha^{(1)}\tilde{\chi}_\alpha^{(0)}+\hat{p}^\mu \kappa_\mu \tilde{\chi}_\alpha^{(0)}.  
\end{align}
According to the fundamental theory of degenerate perturbation, when the inhomogeneity is orthogonal to the solution of the associated homogeneous equation Eq.(\ref{unpert}), i.e.,
\begin{align}
\label{soluble}
(\tilde{\psi}_\gamma,\,\tau\omega^{(1)}\tilde{\chi}^{(0)}+\hat{p}^\mu \kappa_\mu \tilde{\chi}^{(0)})=0,\quad \gamma=1\cdots, 11
\end{align}
with 
\begin{align}
\label{line}
\tilde{\chi}_\alpha^{(0)}=\sum_{i=1}^{11}C_{\alpha\beta}\tilde{\psi}_\beta,
\end{align}
 a unique solution to Eq.(\ref{pert1}) exists. To ensure the existence of nontrivial solutions of Eq.(\ref{soluble}), the frequency $\omega$ has to obey the dispersion relation, i.e., the secular equation,
 \begin{align}
 \text{Det}W_{\gamma\beta}=0,
 \end{align}
 with the matrix elements taking the form,
 \begin{align}
 W_{\gamma\beta}=\omega^{(1)}(\tilde{\psi}_\gamma,\tau\tilde{\psi}_\beta)+ (\tilde{\psi}_\gamma,\hat{p}\cdot\kappa\tilde{\psi}_\beta).
 \end{align}

 With all these matrix elements calculated in Appendix.(\ref{pertu}), the secular equation is,
\begin{align}
\label{secu}
& 
\left|\begin{array}{cccccccccccc}
\omega^{(1)}&0&H_{1,3}&0&0&0  & 0 & 0&0&0&0\\
 0&\omega^{(1)}&H_{2,3}&0&0&0 & 0& 0&0&0&0\\
 H_{1,3}&H_{2,3}&\omega^{(1)}&0&0&0& 0 &  0&0&0&0\\
 0&0&0&\omega^{(1)}&0&0  &0 & 0  &0&0&0\\
 0&0&0&0&\omega^{(1)}&0  &0 & 0  &0&0&0\\
 0&0&0&0&0&\omega^{(1)}  & 0 & 0&0&0&0\\
 0&0&0&0&0&0 & \omega^{(1)}& 0&H_{7,9}&0&0\\
 0&0&0&0&0&0& 0 &  \omega^{(1)}&0&H_{7,9}&0\\
 0&0&0&0&0&0  &H_{7,9} & 0  &\omega^{(1)}&0&0\\
 0&0&0&0&0&0  &0 & H_{7,9}  &0&\omega^{(1)}&0\\
 0&0&0&0&0&0  &0 & 0  &0&0&\omega^{(1)}\\
\end{array}\right|=0 \,,
\end{align}
the roots of this equation are
\begin{align}
&\omega^{(1)}_1=-\omega^{(1)}_2=\sqrt{H^2_{2,3}+H^2_{1,3}}{},\quad \omega^{(1)}_{3}=\omega^{(1)}_{4}=\omega^{(1)}_{5}=0,\nn\\
&\omega^{(1)}_6=\omega^{(1)}_{11}=0,\quad\omega^{(1)}_7=\omega^{(1)}_8=H_{7,9},\quad \omega^{(1)}_{9}=\omega^{(1)}_{10}=-H_{7,9},
\end{align}
and one appropriate linear combination of $\tilde{\psi}$ satisfying the solubility condition Eq.(\ref{soluble}) can be chosen to be
\begin{align}
&\tilde{\chi}^{(0)}_1=\frac{1}{\sqrt{2}}(-\frac{H_{1,3}}{\sqrt{H^2_{1,3}+H^2_{2,3}}}\tilde{\psi}_1-\frac{H_{2,3}}{\sqrt{H^2_{1,3}+H^2_{2,3}}}\tilde{\psi}_2+\tilde{\psi}_3),\quad\tilde{\chi}^{(0)}_2=\frac{1}{\sqrt{2}}(\frac{H_{1,3}}{\sqrt{H^2_{1,3}+H^2_{2,3}}}\tilde{\psi}_1+\frac{H_{2,3}}{\sqrt{H^2_{1,3}+H^2_{2,3}}}\tilde{\psi}_2+\tilde{\psi}_3),\nn\\
&\tilde{\chi}^{(0)}_3=\frac{1}{\sqrt{\frac{H_{2,3}^2}{H^2_{1,3}}+1}}(-\frac{H_{2,3}}{H_{1,3}}\tilde{\psi}_1+\tilde{\psi}_2),\quad \tilde{\chi}^{(0)}_4=\tilde{\psi}_4,\quad \tilde{\chi}^{(0)}_5=\tilde{\psi}_5,\quad \tilde{\chi}^{(0)}_6=\tilde{\psi}_6,\quad\nn\\
& 
\tilde{\chi}^{(0)}_{7}=\frac{1}{\sqrt{2}}(\tilde{\psi}_7-\tilde{\psi}_9),\quad  \tilde{\chi}^{(0)}_{8}=\frac{1}{\sqrt{2}}(\tilde{\psi}_8-\tilde{\psi}_{10}),\quad\tilde{\chi}^{(0)}_{9}=\frac{1}{\sqrt{2}}(\tilde{\psi}_7+\tilde{\psi}_9),\quad  \tilde{\chi}^{(0)}_{10}=\frac{1}{\sqrt{2}}(\tilde{\psi}_8+\tilde{\psi}_{10}),\quad \tilde{\chi}^{(0)}_{11}=\tilde{\psi}_{11}.
\end{align}
One can readily verify that the results of the first five modes are the same as those in \cite{DeGroot:1980dk}, which is independent of the details of interactions involved. It is reasonable because we are solving the same problem as far as the spinless modes are concerned.
  By substituting the zeroth order eigenvalues and eigenfunctions into Eq.(\ref{pert1}) the first-order eigenfunctions are 
 \begin{align}
 \label{chi1}
 \tilde{\chi}_\alpha^{(1)}=\frac{i}{\gamma\tau}(\tau\omega_\alpha^{(1)}\tilde{\chi}_\alpha^{(0)}+\hat{p}^\mu \kappa_\mu \tilde{\chi}_\alpha^{(0)}),\quad \alpha=1,2\cdots ,11.
 \end{align}

 Subsequently, the second order perturbation equation reads,
\begin{align}
\label{pert2}
&-iL_1\tilde{\chi}^{(2)}= \tau\omega^{(1)}\tilde{\chi}^{(1)}+\hat{p}^\mu \kappa_\mu \tilde{\chi}^{(1)}+\tau\omega^{(2)}\tilde{\chi}^{(0)}.
\end{align}
At this time, the solubility condition turns into
\begin{align}
\label{soluble2}
(\tilde{\chi}^{(0)}_\alpha,\,\tau\omega_\beta^{(1)}\tilde{\chi}^{(1)}_\beta+\hat{p}^\mu \kappa_\mu \tilde{\chi}^{(1)}_\beta+\tau\omega_\beta^{(2)}\tilde{\chi}_\beta^{(0)})=0,
\end{align}
equivalently, this can be written as with the assistance of bracket notation,
\begin{align}
\label{bracket}
\omega^{(2)}_\alpha=i[\tilde{\chi}^{(1)}_\alpha,\tilde{\chi}^{(1)}_\alpha],\quad \alpha=1,2\cdots ,11,
\end{align}
where $\tilde{\chi}^{(1)}$ is given by Eq.(\ref{chi1})  and the bracket is defined as  $[B,C]\equiv (L_1[B],C)$. Substitute the mutilated operator $L_1$ into Eq.(\ref{bracket}), and we get
 \begin{align}
 \label{bzd}
 \omega^{(2)}_\alpha=&-i\big(  (\hat{p}^\mu \kappa_\mu+\tau\omega^{(1)}_\alpha )\tilde{\chi}^{(0)}_\alpha,\, \frac{1}{\gamma\tau}(\hat{p}^\mu \kappa_\mu+\tau\omega^{(1)}_\alpha )\tilde{\chi}^{(0)}_\alpha\big)\nn\\
 =&-\frac{i}{\gamma}\big(\,(\tilde{\chi}^{(0)}_\alpha\tilde{\chi}^{(0)}_\alpha,\frac{1}{\tau}\hat{p}^\mu \kappa_\mu\hat{p}^\nu \kappa_\nu)+\omega^{(1)}_\alpha\omega^{(1)}_\alpha(\tilde{\chi}^{(0)}_\alpha,\tau\tilde{\chi}^{(0)}_\alpha)+2\omega^{(1)}_\alpha(\tilde{\chi}^{(0)}_\alpha,\hat{p}^\mu \kappa_\mu\tilde{\chi}^{(0)}_\alpha)\big).
 \end{align}
  With a rather lengthy calculation, we obtain the frequencies in terms of various thermodynamic integrals $I, L$ and $N$ displayed in Appendix.(\ref{int})
 \begin{align}
 \label{frequency}
& \omega_1=\sqrt{H^2_{2,3}+H^2_{1,3}}-\frac{i}{\gamma}(Q_{1,1}-H^2_{2,3}-H^2_{1,3}),\nn\\ &\omega_2=-\sqrt{H^2_{2,3}+H^2_{1,3}}-\frac{i}{\gamma}(Q_{1,1}-H^2_{2,3}-H^2_{1,3}),\quad \nn\\
&  \omega_3=-\frac{i}{\gamma}Q_{3,3},\quad  \omega_4=-\frac{i}{\gamma}Q_{4,4},\quad
  \omega_5=-\frac{i}{\gamma}Q_{4,4},\quad 
  \omega_6=-\frac{i}{\gamma}Q_{6,6},\quad\nn\\
  &\omega_7=H_{7,9}-\frac{i}{\gamma}(Q_{7,7}-H^2_{7,9}),\quad \omega_8=H_{7,9}-\frac{i}{\gamma}(Q_{7,7}-H^2_{7,9}),\quad \nn\\ &\omega_9=-H_{7,9}-\frac{i}{\gamma}(Q_{9,9}-H^2_{7,9}),\quad  \omega_{10}=-H_{7,9}-\frac{i}{\gamma}(Q_{9,9}-H^2_{7,9}),\quad \omega_{11}=-\frac{i}{\gamma}Q_{11,11},
 \end{align}
where the matrix elements $Q_{i,j}$ and $H_{i,j}$ are calculated and given in Appendix.(\ref{pertu}) and (\ref{pertu1}). By comparison with the results in \cite{DeGroot:1980dk}, one can derive the detailed results for all first-order transport coefficients of ordinary hydrodynamics,  but we here only concentrate on the spin modes. Here we comment that these spin-related dispersion relations have no match with the remaining six ones in \cite{Hattori:2019lfp}, because in that work the linear analysis concentrates on the non-conservative spin density.  Among the spin modes there are four propagating modes, two degenerate modes with the propagating speed $H_{7,9}/\kappa$ damp according to the damping rate  $-[\tilde{\chi}^{(1)}_7,\tilde{\chi}^{(1)}_7]$  while the other two travel in the opposite direction with the damping rate $-[\tilde{\chi}^{(1)}_9,\tilde{\chi}^{(1)}_9]$. On the other hand, the sixth and eleventh modes are purely decaying at their respective decaying rates.

It is interesting to note that the imaginary parts of these frequencies encode the information of the relaxation of related dissipative quantities, i.e, the dissipation of a perturbation of conserved charges around the equilibrium state. If the system of our interest is not too far away from the equilibrium state, in that case the linear analysis suffice and we can get a quantitative comparison of two typical relaxation times, one is the relaxation time for energy momentum tensor, the other is for spin tensor, which is significantly crucial in the investigation of spin polarization. Before that, we need to clarify the definition of the spin relaxation time. As is shown in \cite{Hu:2022lpi}, we propose that the relaxation time for spin density should be taken as the longest one of all relaxation times for six spin modes,
\begin{align}
\tau_s=\max\{\frac{1}{|\omega_\alpha^{(2)}|},\alpha=6,\cdots 11\}
\end{align}
with the sign $|A|$ representing  the amplitude of a complex $A$. To clarify the equilibrium picture, we compare it with  another typical time scale, the relaxation time for shear modes $\omega_{4,5}$, which describes the dissipation of an initial disturbance through shear viscosity $\eta$. We denote it with $\tau_{\eta}$ to reflect that this time scale is closely connected with the relaxation of transverse momentum density via shear viscosity $\eta$. If the former is far smaller than the latter,  the initial disturbance has minor effects on spin evolution because the spin density relaxes monotonically and quickly to the equilibrium much earlier than other spinless modes and almost independent. Reversely, the non-equilibrium effects brought by the spin degrees of freedom play a big role and may be irrelevant to heavy-ion collisions for much longer spin equilibration time. In an intermediate case where these two scales are comparable, the status of spin hydrodynamics is enhanced because this is exactly within the range of the application for spin hydrodynamic theory.

\begin{figure}[!htb]
	\includegraphics[width=0.4\textwidth]{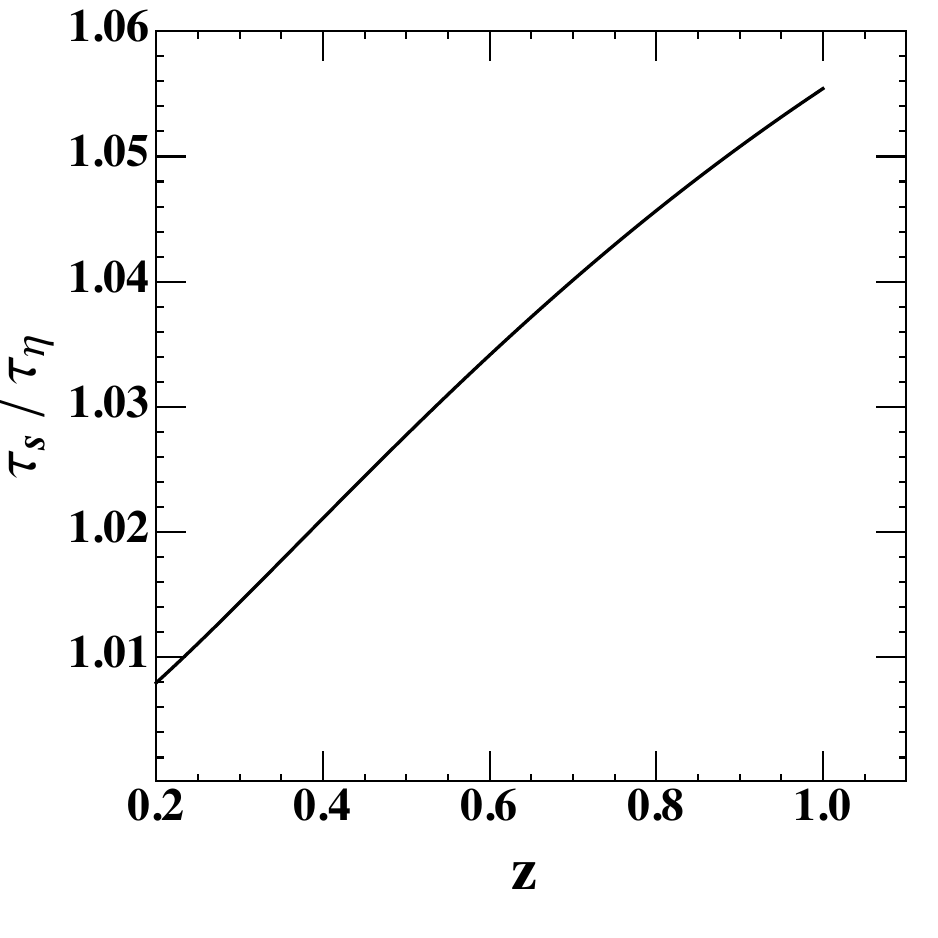} \ ~ \ ~  
	\caption{The relative relaxation time defined by the ratio of two distinct typical time scales $\tau_s/\tau_{\eta}$ as a function of $z(= \frac{m}{T})$.}
	\label{fig1}
\end{figure}

To that end, we need to specify the parameters in our calculation. The spin Boltzmann equation is specially derived for the spin transport for massive quarks, i.e., strange quarks ($m=150 \,\text{Mev}$) temporarily overlooking the interactions with massless quarks and gluons.  The relevant temperature  of the hot matter created in heavy-ion collisions ranges from about $T_c$ to $5T_c$ (the critical temperature $T_c$ is $170\, \text{Mev}$). Since not all thermodynamic integrals can be evaluated analytically, we calculate the frequencies of normal modes numerically. The dependence of the relative relaxation time on the reduced mass $z$ is exhibited in Fig.\ref{fig1}.
From this figure, we can see it clearly that the spin relaxation time is a little larger than shear relaxation time and their ratio grow with the increasing reduced mass. Throughout  temperature region we are interested in, their departure from each other is rather small. Therefore, these two typical times are comparable satisfying the intermediate scenario introduced in the proceeding paragraphs, which calls for the tangled dynamic evolution of both spin and momentum (their characteristic equilibration times are almost the same), while identical conclusions are also reported in related works recently \cite{Ambrus:2022yzz,Weickgenannt:2022zxs}. It seems that increasing temperature will lessen the separation of two time scales. In the hot medium where strange quarks are generated, spin and moment relax to equilibrium via the collisions between strange quarks at almost the same time. In a different viewpoint, the dependence of $\tau_s/\tau_{\eta}$ on $z$ can be also interpreted by holding $T$ fixed, which reads that increase particle mass enhances the effect of separation for two equilibration time scales. The slowness of spin dissipation may be attributed to suppression of spin interaction by the mass of constitute particles, which is consistent with the reason why spin rotation is suppressed in microscopic collision process proposed in \cite{Hattori:2019lfp}. Nevertheless, present results has dependence on the parameterization of $\gamma$. For the impacts of the energy dependence of $\gamma$ on the separation of two time scales discussed here, we leave it to a future work \cite{Hu:2022mvl}.

\clearpage

 \section{Summary and outlook}
 \label{su}
 In this work, we present a detailed linear analysis of normal modes of linearized collision operator based on the spin Boltzmann equation for massive fermions proposed in \cite{Weickgenannt:2021cuo}. When neglecting the non-diagonal part of the transition rate, the linearized collision operator $L_1$ is proved to be Hermite and solving normal modes can be done in the fashion of degenerate perturbation theory.
 Moreover, the reformed collision kernel phenomenologically incorporate the important ansatz of the conservation of total angular momentum in a collision event. The  eleven zero modes exactly correspond to eleven conserved charges that are tightly associated with spin hydrodynamics. Considering the complexity of solving integral equation, we instead approximate the full linearized collision operator with a mutilated operator, which inherits necessary properties of linearized collision operator. 
 
 With the simplified collision kernel, solving normal modes completely reduces to numerical integration order by order and we 
 calculate the frequencies of these normal modes growing from zero modes up to second order in wave vector. We also show that our framework can be well applied to investigate the relaxation of spin. Identifying spin equilibration time as the largest one of all reciprocals  of damping rates for spin modes, we compare it with typical time scale for momentum equilibration. Our results manifest that these two time scales are close over the temperature range of our interest, which also show that increasing particle mass will lead to separation of scales while high temperature puts them closer. Therefore, the evolution of spin can't be independent of that of momentum treating it as equilibrated. The clarification of the hierarchy for relaxation times based on reliable quantum kinetic theory is highly non-trivial in resolving  the problem of discovering the local spin polarization in the experiments of relativistic heavy-ion collisions.
 
  There are still some improvements or possible extensions to our evaluation. Firstly, the parameter $\gamma$ introduced to represent the eigenvalue with infinite degeneracy is set to be spin and momentum independent, otherwise $\gamma$ can't be factorized out of the integral Eq.(\ref{bzd}).  In our framework, $\gamma$ can be naturally parameterized as momentum dependent without contradicting the conservation laws compared to traditional RTA. In principle, $\gamma$ is not a free parameter and supposed to be determined by solving or approximately solving eigen spectrum of the full linearized collision operator, which is left as a further research in future. Secondly, the adopted kinetic equation is derived only for massive quarks, although massive strange quarks should contribute a lot as constitute components of $\Lambda$ hyperons. However there are also other processes contributing to spin and  momentum relaxation such as collisions between strange quarks and massless  gluons and u,d quarks.  They shall play a role but whether those processes are predominant over collisions between massive quarks themselves calls for a further investigation.  For completeness and precision, the scattering of strange quarks and massless quarks and gluons  is also necessarily considered. Last but not the least, we take one of the easiest equilibrium configuration, i.e., $\Omega=0$, on top of which the linear analysis is carried out. It is generally believed that finite thermal vorticity can survive in global equilibrium and so is the spin potential $\Omega$. If choosing the configuration with finite vorticity, we are allowed with another power counting scheme where vorticity field breaks rotation symmetry and the theory  then is anisotropic.
  
\section*{Acknowledgments}

 This work was supported by the NSFC Grant No.11890710, No.11890712 and No.12035006.
\clearpage
\begin{appendix}
	
\section{Thermodynamic  Integral} \label{int}
To proceed, we first introduce the following integration formula
\begin{align}
\label{inq}
I^{(r)}_{\alpha_1\cdots\alpha_n}&\equiv 2\int \frac{\rm dP}{(2\pi)^3 }\,p_{\alpha_1}p_{\alpha_2}\cdots p_{\alpha_n}\frac{e^{\xi-\beta \cdot p}}{(u\cdot p)^r}\nn\\
&=I^{(r)}_{n0}u_{\alpha_1}\cdots u_{\alpha_n}+I^{(r)}_{n1}(\Delta_{\alpha_1\alpha_2}u_{\alpha_3\cdots\alpha_n}+\text{permutations})+\cdots,
\end{align}
where the abbreviation dP stands for $d^4p\, \delta(p^2 - m^2)$ and the formal expression after the second equality comes from the analysis of Lorentz covariance. Using the projection operator $u^\alpha$ and $\Delta^{\alpha\beta}$, the scalar coefficients in the form of thermodynamic integrals are given by 
\begin{eqnarray}
\label{Inq}
I^{(r)}_{nq} &\equiv& \frac{2}{(2q+1)!!} \int \frac{\rm dP}{(2\pi)^3}\,(u\cdot p)^{n-2q-r} (\Delta_{\alpha\beta} p^{\alpha} p^{\beta})^q e^{\xi-\beta \cdot p},
\end{eqnarray}
where $K_n(z)$ denotes the modified Bessel functions of the second kind defined as
\begin{eqnarray}
K_n(z) &\equiv& \int_0^{\infty} \mathrm{d}x\, \cosh(nx)\, e^{- z \cosh x}.
\end{eqnarray}
Specially, we note that $I^{(0)}_{10}=n$, $I^{(0)}_{20}=e,\, I^{(0)}_{21}=-P$ , $I^{(0)}_{31}=-hT$ and 
\begin{align}
I^{(0)}_{30}(z)=\frac{T^5z^5e^{\xi}}{32\pi^2}\big(K_5(z)+K_3(z)-2K_1(z)\big),
\end{align}
with $z\equiv\frac{m}{T}$, $n, e, P, h$ are the number  density, energy density, static pressure and enthalpy respectively.

In addition, when handling the angular momentum integrations, the following similar and useful formulas are also of our interest,
\begin{align}
L^{(r)}_{\alpha_1\cdots\alpha_n}&\equiv 2\int \frac{\rm dP}{(2\pi)^3 (p^0+m)}\,p_{\alpha_1}p_{\alpha_2}\cdots p_{\alpha_n}\frac{e^{\xi-\beta \cdot p}}{(u\cdot p)^r}\nn\\
&=L^{(r)}_{n0}u_{\alpha_1}\cdots u_{\alpha_n}+L^{(r)}_{n1}(\Delta_{\alpha_1\alpha_2}u_{\alpha_3\cdots\alpha_n}+\text{permutations})+\cdots,\nn\\
N^{(r)}_{\alpha_1\cdots\alpha_n}&\equiv 2\int \frac{\rm dP}{(2\pi)^3 (p^0+m)^2}\,p_{\alpha_1}p_{\alpha_2}\cdots p_{\alpha_n}\frac{e^{\xi-\beta \cdot p}}{(u\cdot p)^r}\nn\\
&=N^{(r)}_{n0}u_{\alpha_1}\cdots u_{\alpha_n}+N^{(r)}_{n1}(\Delta_{\alpha_1\alpha_2}u_{\alpha_3\cdots\alpha_n}+\text{permutations})+\cdots.
\end{align}
The scalar functions $L^{(r)}_{nq}$ and $N^{(r)}_{nq}$ can be also defined like Eq.(\ref{Inq}). With extra factor appearing in the integrations, these integrals can not be expressed with the modified Bessel functions of the second kind $K_n(z)$.

\section{Normalized factors} \label{int1}
First, we define two auxiliary unit vectors $j^\mu$ and $v^\mu$, which satisfy
\begin{align}
&u\cdot l=u\cdot j=u\cdot v=l\cdot j=l\cdot v=j\cdot v=0,\nn\\
&l^2=j^2=v^2=-1.
\end{align}
Thus we can expand $p^\mu$ and $J^{\mu\nu}$ as
\begin{align}
&p^\mu=u\cdot p \,u^\mu+l\cdot p \,l^\mu+j\cdot p \,j^\mu+v\cdot p\, v^\mu,\nn\\
&J^{\mu\nu}=u_\mu J^{\mu\nu}l_\nu-l_\mu J^{\mu\nu}u_\nu+u_\mu J^{\mu\nu}j_\nu-j_\mu J^{\mu\nu}u_\nu+u_\mu J^{\mu\nu}v_\nu-v_\mu J^{\mu\nu}u_\nu\nn\\
&\quad\,\,\,\,+l_\mu J^{\mu\nu}j_\nu-j_\mu J^{\mu\nu}l_\nu+l_\mu J^{\mu\nu}v_\nu-v_\mu J^{\mu\nu}l_\nu+j_\mu J^{\mu\nu}v_\nu-v_\mu J^{\mu\nu}j_\nu.
\end{align}
Accounting for the antisymmetric property of total angular momentum $J^{\mu\nu}$, the effective degrees of freedom, or the effective basis can be chosen as $(1,u\cdot p,l\cdot p,j\cdot p,v\cdot p, u_\mu J^{\mu\nu}l_\nu,u_\mu J^{\mu\nu}j_\nu,u_\mu J^{\mu\nu}v_\nu,l_\mu J^{\mu\nu}j_\nu,l_\mu J^{\mu\nu}v_\nu,j_\mu J^{\mu\nu}v_\nu)$, and they are label by the $i$-th basis respectively ($i=1,2,\cdots,11$).

The normalized factors for the zeroth eigenfunctions meeting the condition of Eq.(\ref{one}) are 
\begin{eqnarray}
\label{Inq3}
V_{1,1}&=&\exp(\xi)\int \frac{\rm d\Gamma}{(2\pi)^3}\frac{u\cdot p}{T}\exp(-\beta \cdot p)=\frac{n}{T},\nn\\
\quad V_{2,2}&=&\exp(\xi)\int \frac{\rm d\Gamma}{(2\pi)^3}\frac{(u\cdot p-\frac{e}{n})^2(u\cdot p)}{T^3}\exp(-\beta \cdot p)=\frac{I^{(0)}_{30}-\frac{e^2}{n}}{T^3},\nn\\
V_{3,3}&=&V_{4,4}=V_{5,5}=\exp(\xi)\int \frac{\rm d\Gamma}{(2\pi)^3}\frac{(u\cdot p)(l\cdot p)^2}{T^3}\exp(-\beta \cdot p)=\frac{h}{T^2}.
\end{eqnarray}

Because of the calibration settings mentioned before, we have $\hat{t}=u=(1,0,0,0)$, then the other normalized factors are
\begin{align}
V_{6,6}=V_{7,7}=V_{8,8}&=\exp(\xi)\int \frac{\rm d\Gamma}{(2\pi)^3}\frac{u\cdot p}{T}u_\mu J^{\mu\nu}l_\nu u_\rho J^{\rho\sigma}l_\sigma\exp(-\beta \cdot p)\nn\\
&=\frac{1}{2m^2T}(-I^{(0)}_{31}+2L^{(0)}_{41}-N^{(0)}_{51}),\nn\\
V_{9,9}=V_{10,10}=V_{11,11}&=\exp(\xi)\int \frac{\rm d\Gamma}{(2\pi)^3}l_\mu J^{\mu\nu}j_\nu(p\cdot u) l_\rho J^{\rho\sigma}j_\sigma\exp(-\beta \cdot p)\nn\\
&=\frac{I^{(0)}_{30}+I^{(0)}_{31}+4L^{(0)}_{41}+10N^{(0)}_{52}}{4m^2T},
\end{align}
where the thermodynamic integrals are given in Appendix.(\ref{int}). We have checked that the factors $V_{2,2},V_{6,6}$ and $V_{9,9}$ are all positive.

\section{Perturbation Matrix Elements I } \label{pertu}
In this section, we calculate the perturbation matrix elements which partly breaks the degeneracy of normal modes. The matrix elements are given by $H_{i,j}=\frac{\exp(\xi)}{T}\int \frac{\rm d\Gamma}{(2\pi)^3}\tilde{\psi}_i\kappa\cdot p\,\tilde{\psi}_j\exp(-\beta \cdot p)$. When indices $i$ and $j$ range from $1$ to $5$, only the following  matrix elements are nonzero.
\begin{eqnarray}
\label{pk}
H_{1,3}&=&\frac{\exp(\xi)}{\sqrt{V_{1,1}V_{3,3}}T^2}\int \frac{\rm d\Gamma}{(2\pi)^3}\kappa\cdot p\,l\cdot p\exp(-\beta \cdot p)=\frac{P\kappa}{\sqrt{V_{1,1}V_{3,3}}T^2},\nn\\
H_{2,3}&=&\frac{\exp(\xi)}{\sqrt{V_{2,2}V_{3,3}}T^3}\int \frac{\rm d\Gamma}{(2\pi)^3}(u\cdot p-\frac{e}{n})p\cdot\kappa(l\cdot p)\exp(-\beta \cdot p)=\frac{P\kappa}{\sqrt{V_{2,2}V_{3,3}}T^2},
\end{eqnarray}
which break the five-fold degeneracy into three-fold. When it comes to spin modes, we first introduce a  four-indice tensor $H^{\mu\nu\rho\sigma}$, which is defined as
\begin{align}
&H^{\mu\nu\rho\sigma}=\frac{\exp(\xi)}{T}\int \frac{\rm d\Gamma}{(2\pi)^3}J^{\mu\nu}(p\cdot \kappa)J^{\rho\sigma}\exp(-\beta \cdot p)\nn\\
&\quad\,\,\,\,=\frac{1}{4m^2}(g^{\mu\sigma}(\kappa^\nu u^\rho+\kappa^\rho u^\nu)I^{(0)}_{31}+g^{\nu\rho}(\kappa^\mu u^\sigma+\kappa^\sigma u^\mu)I^{(0)}_{31}-g^{\nu\sigma}(\kappa^\mu u^\rho+\kappa^\rho u^\mu)I^{(0)}_{31}-g^{\mu\rho}(\kappa^\nu u^\sigma+\kappa^\sigma u^\nu)I^{(0)}_{31})\,)\nn\\
&\quad\,\,\,\,+\frac{1}{4m^2}(L^{(0)}_{41}g^{\mu\rho}\kappa^\sigma u^\nu-5L^{(0)}_{42}g^{\mu\rho}u^\sigma\kappa^\nu-L^{(0)}_{41}g^{\mu\sigma}\kappa^\rho u^\nu+5L^{(0)}_{42}g^{\mu\sigma}u^\rho\kappa^\nu)\nn\\
&\quad\,\,\,\,-\frac{1}{4m^2}(L^{(0)}_{41}g^{\nu\rho}\kappa^\sigma u^\mu-5L^{(0)}_{42}g^{\nu\rho}u^\sigma\kappa^\mu-L^{(0)}_{41}g^{\nu\sigma}\kappa^\rho u^\mu+5L^{(0)}_{42}g^{\nu\sigma}u^\rho\kappa^\mu)\nn\\
&\quad\,\,\,\, +\frac{1}{4m^2}(L^{(0)}_{41}g^{\rho\mu}\kappa^\nu u^\sigma-5L^{(0)}_{42}g^{\rho\mu}u^\nu\kappa^\sigma-L^{(0)}_{41}g^{\rho\nu}\kappa^\mu u^\sigma+5L^{(0)}_{42}g^{\rho\nu}u^\mu\kappa^\sigma)\nn\\
&\quad\,\,\,\,-\frac{1}{4m^2}(L^{(0)}_{41}g^{\sigma\mu}\kappa^\nu u^\rho-5L^{(0)}_{42}g^{\sigma\mu}u^\nu\kappa^\rho-L^{(0)}_{41}g^{\sigma\nu}\kappa^\mu u^\rho+5L^{(0)}_{42}g^{\sigma\nu}u^\mu\kappa^\rho)\nn\\
&\quad\,\,\,\,+\frac{1}{4m^2}[g^{\mu\rho}\big(N^{(0)}_{51}(\kappa^{\nu}u^{\sigma}+u^\nu \kappa^\sigma)+5N^{(0)}_{52}(u^\nu\kappa^\sigma+u^\sigma\kappa^\nu\,)\,\big)-g^{\nu\rho}\big(N^{(0)}_{51}(\kappa^{\mu}u^{\sigma}+u^\mu \kappa^\sigma)+5N^{(0)}_{52}(u^\mu\kappa^\sigma+u^\sigma\kappa^\mu\,)\,\big)\nn\\
&\quad\quad\quad\;-(\rho\leftrightarrow\sigma)]\nn\\
&\quad\,\,\,\,-\frac{1}{4m^2}[N^{(0)}_{51}g^{\mu\rho}(\kappa^\nu u^\sigma+\kappa^\sigma u^\nu\,)-N^{(0)}_{51}g^{\nu\rho}(\kappa^\mu u^\sigma+\kappa^\sigma u^\mu)-(\rho\leftrightarrow\sigma)],
\end{align}
by projecting $H^{\mu\nu\rho\sigma}$ with four direction vectors $(u,l,j,v)$, we get the following non-vanishing matrix elements,
\begin{eqnarray}
H_{7,9}=H_{8,10}&=&\frac{\exp(\xi)}{\sqrt{V_{6,6}V_{9,9}}}\int \frac{\rm d\Gamma}{(2\pi)^3}\frac{p\cdot\kappa}{T}u_\mu J^{\mu\nu}j_\nu l_\rho J^{\rho\sigma}j_\sigma\exp(-\beta \cdot p)=\frac{(-I^{(0)}_{31}+L^{(0)}_{41}-5L^{(0)}_{42}+5N^{(0)}_{52})\kappa}{4m^2T\sqrt{V_{6,6}V_{9,9}}}.
\end{eqnarray}
It is easy to find that the perturbation matrix has a symmetry of transposition. To be concrete, the matrix elements satisfy $H_{i,j}=H_{j,i}$. We now see that there are no cross terms between  the spinless part and the spin part in the whole matrix $W$, which is clear from Eq.(\ref{secu}).
\clearpage
\section{Perturbation Matrix Elements II} \label{pertu1}
In this section, we calculate the 2nd-order perturbation matrix elements appearing in Eq.(\ref{frequency}). The needed integrals are implemented with the assistance of Eq.(\ref{inq}),
\begin{eqnarray}
\label{pk1}
q_{1,1}&=&\frac{\exp(\xi)}{V_{1,1}T}\int \frac{\rm d\Gamma}{(2\pi)^3}\frac{(p\cdot \kappa)^2}{p\cdot u}\exp(-\beta \cdot p)=-\frac{I^{(1)}_{21}}{V_{1,1}T}\kappa^2,\nn\\
q_{2,2}&=&\frac{\exp(\xi)}{V_{2,2}T^3}\int \frac{\rm d\Gamma}{(2\pi)^3}(u\cdot p-\frac{e}{n})\frac{(p\cdot\kappa)^2}{u\cdot p}(u\cdot p-\frac{e}{n})\exp(-\beta \cdot p)=(-I^{(0)}_{31}-\frac{e^2I^{(1)}_{21}}{n^2}+\frac{2eI^{(0)}_{21}}{n})\frac{\kappa^2}{V_{2,2}T^3},\nn\\
q_{1,2}&=&\frac{\exp(\xi)}{\sqrt{V_{1,1}V_{2,2}}T^2}\int \frac{\rm d\Gamma}{(2\pi)^3}\frac{(p\cdot\kappa)^2}{u\cdot p}(u\cdot p-\frac{e}{n})\exp(-\beta \cdot p)=(-I^{(0)}_{21}+\frac{eI^{(1)}_{21}}{n})\frac{\kappa^2}{\sqrt{V_{1,1}V_{2,2}}T^2},\nn\\
q_{3,3}&=&\frac{\exp(\xi)}{V_{3,3}T^3}\int \frac{\rm d\Gamma}{(2\pi)^3}\frac{(p\cdot \kappa)^2(p\cdot l)^2}{p\cdot u}\exp(-\beta \cdot p)=\frac{3I^{(1)}_{42}}{V_{3,3}T^3}\kappa^2,\nn\\
\end{eqnarray}
and the final results for spinless modes are
\begin{eqnarray}
Q_{1,1}&=&Q_{2,2}=\frac{H_{1,3}^2}{2(H^2_{1,3}+H^2_{2,3})}q_{1,1}+\frac{H_{2,3}^2}{2(H^2_{1,3}+H^2_{2,3})}q_{2,2}+\frac{1}{2}q_{3,3}+\frac{H_{1,3}H_{2,3}}{H^2_{1,3}+H^2_{2,3}}q_{1,2},\nn\\
Q_{3,3}&=&\frac{1}{\frac{H_{2,3}^2}{H^2_{1,3}}+1}(\frac{H_{2,3}^2}{H^2_{1,3}}q_{1,1}+q_{2,2}-\frac{2H_{2,3}}{H_{1,3}}q_{1,2}),\nn\\
Q_{4,4}&=&Q_{5,5}=\frac{\exp(\xi)}{V_{3,3}T^3}\int \frac{\rm d\Gamma}{(2\pi)^3}\frac{(p\cdot \kappa)^2(p\cdot j)^2}{p\cdot u}\exp(-\beta \cdot p)=\frac{I^{(1)}_{42}}{V_{3,3}T^3}\kappa^2.
\end{eqnarray}
It is rather cumbersome to carry out the rest of integrals associated with spin modes. First, we introduce a new four-indice tensor $Q^{\mu\nu\rho\sigma}$,
\begin{align}
&Q^{\mu\nu\rho\sigma}=\frac{\exp(\xi)}{T}\int \frac{\rm d\Gamma}{(2\pi)^3}J^{\mu\nu}\frac{(p\cdot \kappa)^2}{p\cdot u}J^{\rho\sigma}\exp(-\beta \cdot p)\nn\\
&\quad\quad\,\,\,\,=\frac{1}{4m^2T}(-g^{\mu\rho}g^{\nu\sigma}(I^{(1)}_{41}+5I^{(1)}_{42})+g^{\mu\sigma}g^{\nu\rho}(I^{(1)}_{41}+5I^{(1)}_{42})\,)\kappa^2\nn\\
&\quad\quad\,\,\,\,+\frac{1}{4m^2T}\big(-g^{\mu\sigma}u^\nu u^\rho I^{(1)}_{41}-g^{\mu\sigma}\Delta^{\nu\rho} I^{(1)}_{42}+2g^{\mu\sigma}l^{\nu}l^{\rho} I^{(1)}_{42}-g^{\nu\rho}u^\mu u^\sigma I^{(1)}_{41}-g^{\nu\rho}\Delta^{\mu\sigma} I^{(1)}_{42}+2g^{\nu\rho}l^{\mu}l^{\sigma} I^{(1)}_{42}\nn\\
&\quad\quad\quad\quad\,\;\;+\;g^{\nu\sigma}u^\mu u^\rho I^{(1)}_{41}+g^{\nu\sigma}\Delta^{\mu\rho} I^{(1)}_{42}-2g^{\nu\sigma}l^{\mu}l^{\rho} I^{(1)}_{42}+g^{\mu\rho}u^\nu u^\sigma I^{(1)}_{41}+g^{\mu\rho}\Delta^{\nu\sigma} I^{(1)}_{42}-2g^{\mu\rho}l^{\nu}l^{\sigma} I^{(1)}_{42}\,\big)\kappa^2\nn\\
&\quad\quad\,\,\,\,-\frac{1}{2m^2T}\big(-g^{\mu\sigma}u^\nu u^\rho L^{(0)}_{41}-g^{\mu\sigma}\Delta^{\nu\rho} L^{(0)}_{42}+2g^{\mu\sigma}l^{\nu}l^{\rho} L^{(0)}_{42}-g^{\nu\rho}u^\mu u^\sigma L^{(0)}_{41}-g^{\nu\rho}\Delta^{\mu\sigma} L^{(0)}_{42}+2g^{\nu\rho}l^{\mu}l^{\sigma} L^{(0)}_{42}\nn\\
&\quad\quad\quad\quad\,\;\;+\;g^{\nu\sigma}u^\mu u^\rho L^{(0)}_{41}+g^{\nu\sigma}\Delta^{\mu\rho} L^{(0)}_{42}-2g^{\nu\sigma}l^{\mu}l^{\rho} L^{(0)}_{42}+g^{\mu\rho}u^\nu u^\sigma L^{(0)}_{41}+g^{\mu\rho}\Delta^{\nu\sigma} L^{(0)}_{42}-2g^{\mu\rho}l^{\nu}l^{\sigma} L^{(0)}_{42}\,\big)\kappa^2\nn\\
&\quad\quad\,\,\,\,+\frac{1}{2m^2T}[g^{\mu\rho}u^\sigma u^\nu( L^{(1)}_{51}+5L^{(1)}_{52})-g^{\mu\sigma}u^\rho u^\nu( L^{(1)}_{51}+5L^{(1)}_{52})-g^{\nu\rho}u^\sigma u^\mu( L^{(1)}_{51}+5L^{(1)}_{52})+g^{\nu\sigma}u^\rho u^\mu( L^{(1)}_{51}+5L^{(1)}_{52})\,]\kappa^2\nn\\
&\quad\quad\,\,\,\,-\frac{1}{4m^2T}[-\;g^{\mu\rho}\big(u^{\nu}u^{\sigma}N^{(0)}_{51}+(\Delta^{\nu\sigma}-2l^\sigma l^\nu)N^{(0)}_{52}\,\big)+g^{\nu\rho}\big(u^{\mu}u^{\sigma}N^{(0)}_{51}+(\Delta^{\mu\sigma}-2l^\mu l^\sigma)N^{(0)}_{52}\,\big)\nn\\
&\quad\quad\quad\;\quad+g^{\mu\sigma}\big(u^{\nu}u^{\rho}N^{(0)}_{51}+(\Delta^{\nu\rho}-2l^\rho l^\nu)N^{(0)}_{52}\,\big)-g^{\nu\sigma}\big(u^{\mu}u^{\rho}N^{(0)}_{51}+(\Delta^{\mu\rho}-2l^\mu l^\rho)N^{(0)}_{52}\big)]\kappa^2\nn\\
&\quad\quad\,\,\,\,+\frac{1}{4m^2T}[g^{\mu\rho}\big(-u^\nu u^\sigma N^{(1)}_{61}-(\Delta^{\nu\sigma}+5u^\nu u^\sigma-2l^\nu l^\sigma)N^{(1)}_{62}+(-7\Delta^{\nu\sigma}+14l^\nu l^\sigma)N^{(1)}_{63}\big)\nn\\
&\quad\quad\quad\,\,-g^{\nu\rho}\big(-u^\mu u^\sigma N^{(1)}_{61}-(\Delta^{\mu\sigma}+5u^\mu u^\sigma-2l^\mu l^\sigma)N^{(1)}_{62}+(-7\Delta^{\mu\sigma}+14l^\mu l^\sigma)N^{(1)}_{63}\big)\nn\\
&\quad\quad\quad\,\,-g^{\mu\sigma}\big(-u^\nu u^\rho N^{(1)}_{61}-(\Delta^{\nu\rho}+5u^\nu u^\rho-2l^\nu l^\rho)N^{(1)}_{62}+(-7\Delta^{\nu\rho}+14l^\nu l^\rho)N^{(1)}_{63}\big)\nn\\
&\quad\quad\quad\,\,+g^{\nu\sigma}\big(-u^\mu u^\rho N^{(1)}_{61}-(\Delta^{\mu\rho}+5u^\mu u^\rho-2l^\mu l^\rho)N^{(1)}_{62}+(-7\Delta^{\mu\rho}+14l^\mu l^\rho)N^{(1)}_{63}\big)]\kappa^2\nn\\
&\quad\quad\,\,\,\,-\frac{1}{4m^2T}[u^\mu u^\rho\big(-(\Delta^{\nu\sigma}-2l^\nu l^\sigma)N^{(1)}_{62}+(-7\Delta^{\nu\sigma}+14l^\nu l^\sigma)N^{(1)}_{63}\big)\nn\\
&\quad\quad\quad\,\,-u^\nu u^\rho\big(-(\Delta^{\mu\sigma}-2l^\mu l^\sigma)N^{(1)}_{62}+(-7\Delta^{\mu\sigma}+14l^\mu l^\sigma)N^{(1)}_{63}\big)\nn\\
&\quad\quad\quad\,\,-u^\mu u^\sigma\big(-(\Delta^{\nu\rho}-2l^\nu l^\rho)N^{(1)}_{62}+(-7\Delta^{\nu\rho}+14l^\nu l^\rho)N^{(1)}_{63}\big)\nn\\
&\quad\quad\quad\,\,+u^\nu u^\sigma\big(-(\Delta^{\mu\rho}-2l^\mu l^\rho)N^{(1)}_{62}+(-7\Delta^{\mu\rho}+14l^\mu l^\rho)N^{(1)}_{63}\big)]\kappa^2,
\end{align}
from which one can obtain the following indispensable integrals by projecting onto various directions,
\begin{align}
&q_{7,7}=q_{8,8}=\frac{\exp(\xi)}{V_{6,6}T}\int \frac{\rm d\Gamma}{(2\pi)^3}u_\mu J^{\mu\nu}j_\nu\frac{(p\cdot \kappa)^2}{p\cdot u} u_\rho J^{\rho\sigma}j_\sigma\exp(-\beta \cdot p)\nn\\
&\quad\,\,\,\,\,=\frac{1}{4m^2V_{6,6}T}(4I^{(1)}_{42}+2L^{(0)}_{41}+ 2L^{(0)}_{42}-2L^{(1)}_{51}-10L^{(1)}_{52}-N^{(0)}_{52}-N^{(0)}_{51}+N^{(1)}_{61}+5N^{(1)}_{62})\kappa^2,\nn\\
&q_{9,9}=q_{10,10}=\frac{\exp(\xi)}{V_{9,9}T}\int \frac{\rm d\Gamma}{(2\pi)^3}l_\mu J^{\mu\nu}j_\nu\frac{(p\cdot \kappa)^2}{p\cdot u} l_\rho J^{\rho\sigma}j_\sigma\exp(-\beta \cdot p)\nn\\
&\quad\,\,\,\,\,=\frac{1}{4m^2V_{9,9}T}(-I^{(1)}_{41}-I^{(1)}_{42}-8L^{(0)}_{42}+4N^{(0)}_{52}-4N^{(1)}_{62}-28N^{(1)}_{63}\big)\kappa^2,
\end{align}
and finally,
\begin{align}
&Q_{6,6}=\frac{\exp(\xi)}{V_{6,6}T}\int \frac{\rm d\Gamma}{(2\pi)^3}u_\mu J^{\mu\nu}l_\nu\frac{(p\cdot \kappa)^2}{p\cdot u} u_\rho J^{\rho\sigma}l_\sigma\exp(-\beta \cdot p)\nn\\
&\quad\,\,\,\,\,=\frac{1}{4m^2V_{6,6}T}(2I^{(1)}_{42}+2 L^{(0)}_{41}+6L^{(0)}_{42}-2 L^{(1)}_{51}-10L^{(1)}_{52}\,-N^{(0)}_{51}-3N^{(0)}_{52}+ N^{(1)}_{61}+5 N^{(1)}_{62})\kappa^2,\nn\\
&Q_{7,7}=Q_{8,8}=Q_{9,9}=Q_{10,10}=\frac{1}{2}(q_{7,7}+q_{9,9}),\nn\\
&Q_{11,11}=\frac{\exp(\xi)}{V_{9,9}T}\int \frac{\rm d\Gamma}{(2\pi)^3}j_\mu J^{\mu\nu}v_\nu\frac{(p\cdot \kappa)^2}{p\cdot u} j_\rho J^{\rho\sigma}v_\sigma\exp(-\beta \cdot p)\nn\\
&\quad\,\,\,\,\,=\frac{1}{4m^2V_{9,9}T}(-I^{(1)}_{41}-3I^{(1)}_{42}-4L^{(0)}_{42}+2N^{(0)}_{52}-2N^{(1)}_{62}-14N^{(1)}_{63}\big)\kappa^2,
\end{align}
where the diagonal matrix elements $Q_{i,i},\, i=1\cdots 11$,  are  exactly what we present in Eq.(\ref{frequency}). 
\clearpage

\end{appendix}
\bibliographystyle{apsrev}
\bibliography{spinmode}{}

\end{document}